# Design specifications of the Human Robotic interface for the biomimetic underwater robot "yellow submarine project"

Dr Bheemaiah, Anil K  341/42 17 cross Ideal Homes Township, Rajarajeshwarinagar Bangalore 560098, India Anil.bheemaiah@yahoo.co.in

*Abstract: This paper describes the design of a web based multi agent design for a collision avoidance auto navigation biomimetic submarine for submarine hydroelectricity. The paper describes the nature of the map - topology interface for river bodies and the design of interactive agents for the control of the robotic submarine. The agents are migratory on the web and are designed in XML/html interface with both interactive capabilities and visibility on a map. The paper describes mathematically the user interface and the map definition languages used for the multi agent description.*

## 1 Introduction

Multi agents are a distributed approach to the control of robotics in the existence of a front end to the robot for the control and navigation of the robot and for a better exception handling. Agents are used in many domains and fields and is a mature area of software development, predominately web based and web migrating agents, [.Lyle N. Long et al, 3] describes several autonomous vehicle architectures but none with an XML or html web interface, There is much work on independently developed robotic XML languages, like ROBOTML,[ Maxim Makatchev, 1] and [Tingting Fu et al , 2]
This paper delineates the specification of the design of the yellow submarine project, without the need for agent autonomous control software. The inspiration drawn from interactive video books leads directly to this application in a single - one to many mapping of agents to robotic navigation problems.

## 2 Problem Description and Background

TheYellowsubmarineproject, [http://www.friendsofwildlife.com/yellowsubmarine/specifications of the yellow submarine design.html] utilizes a biomimetic robot in design that is a thin client that has to be navigated and positioned in the most appropriate location in a river map. The interface for this from the agent communication language layer is a web interface that provides a visual tool displaying the location of the submarines on a map of the river. The map is inherited from Google earth and displays in a much higher resolution the topology of the river area.

A drag and click interface is designed to move the robot to the designated coordinates that corresponds to the optimal positioning.

Optimal positioning tools are also part of this interface and the subject of another publication.[http://www.friendsofwildlife.com/yellowsubmarine/design specifications of the optimal flow algorithms in the web interface of the biomimetic underwater robotic yellow submarine project.html] This publication delineates the interface to the robotic language layer of the web agent, the design of the web agent and the exception handler.

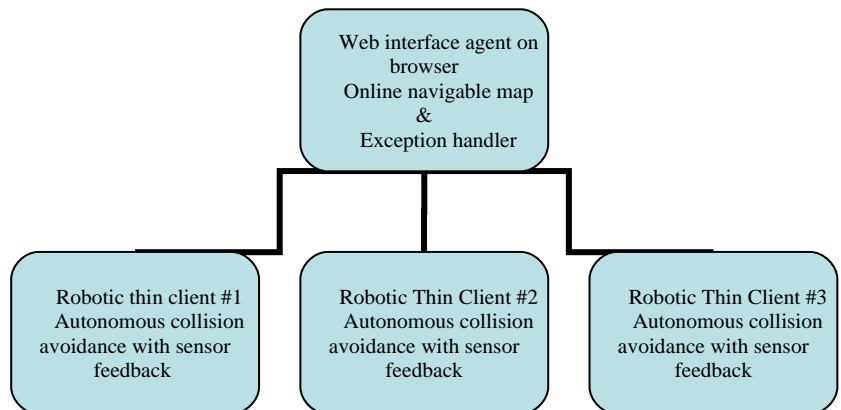

Fig 1: Schematic of the multi agent design

## 2.1 Design of the web agent.

### 2.1.1 The map user interface and click and drag.

The map user interface inherits the Google earth interface for the display of the map of the river in rather high resolution with a symbolic representation of the robot defined by a function call in the html code.

The coordinates of the submarine are returned and polled every 15 seconds by the GPS_get_coordinates instruction. This is displayed on the map. The robot by another function call can for the event of clicking the robot be dragged to the optimal location from where it is moving and parked by using a menu interface.

A map definition language defines the xml markup for the map of the designated robot parking area. The map definition language is a non linear markup indicating in the event:click_on_robot:=get_scale, the scale parameters in the topology that is of significance to the robot parking .

Such as MDL: Coordinates_markers

MDL:Coordinates_landmarks_passed
MDL:Coordinates_lookahead_landmark
MDL:Coordinates_flow_obstacles
MDL:Flow_obstacles

Here the map definition language is a mark up on the map of the flow design algorithms defined in a separate publication for the non linear list of the coordinates of important landmarks, with a look ahead system to aid navigation and location of the robot and the landmarks passed. These land marks are either fed manually on the map or computed from the flow algorithms.

The map definition language also stores the vectors as a linear map and marks it in XML on the map of the flows between the landmarks indicated, thus naturally depending on the scale indicated, there would be two or more landmarks or obstacles per flow listed on the MDL.

2.1.2 The interpreter to the robotic language for this interface.

The interpreter consists of a menu definition with an event language with event definitions.

Event:click_on _robot:=drag robot , place robot.
Event : click_on_robot:=compute optimal flow.

Fi

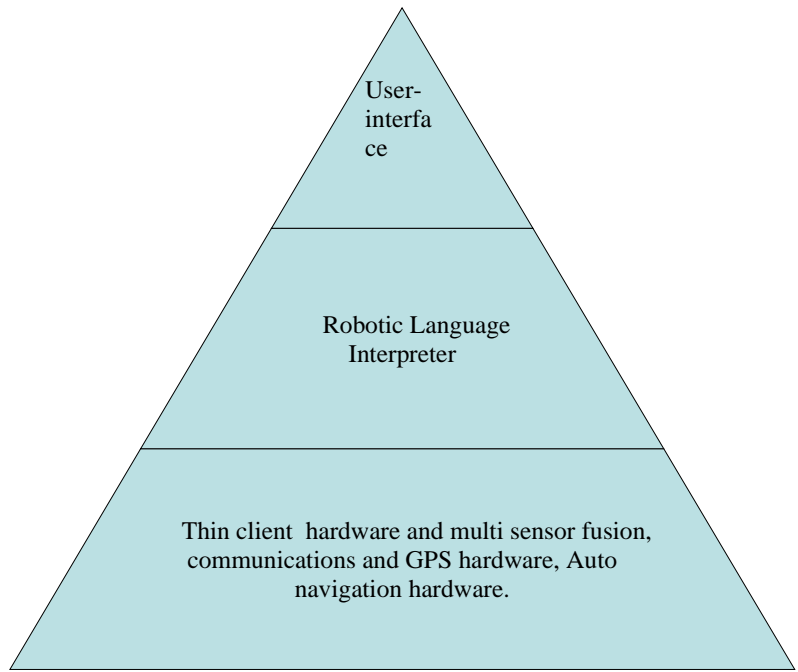

g 2 : Design of the three layers of the agent communication model

### 2.1.3 The exception handler.

Exception handling window in the user interface, defined in xml/html in the design of exceptions.
1. Exp_anchor: in case of exp_communication_failure, exp_GPS _failure or exp_sensor/ power failure,exp_ propulsión _failure.
2. Exp: Auto_park: In case of prolonged timeout, auto park in fuel rendezvous terminal
3. Exp_propulsion: in case the robot is unable to anchor.

## 3  Conclusion and future work

This paper is the design of a simple user interface for the positioning and monitoring of submarine robots. It uses the standard web interface and event language for the user interface with and added interpreter for a defined robotic language for the robot that is defined in another publication: Robotic language specifications for the yellow submarine, biomimetic underwater robot alternative hydro electricity project.

Future work includes a multi agent interface, improvement of the exception handling and the use of a natural language interface for the robot.